\documentclass[preprint]{aastex6}

\begin{document}

\defcitealias{Rigault2013}{R13}
\defcitealias{Rigault2015}{R15}
\defcitealias{Jones2015}{J15}
\newcommand{\mstellar}{\ensuremath{M_{\mathrm{stellar}}}}
\newcommand{\msun}{\ensuremath{M_{\sun}}}
\newcommand{\probpassive}{\ensuremath{\cal P(\mathrm{Ia}\epsilon)}}

\title{Environmental Dependence of Type Ia Supernova Luminosities from a Sample without a Local--Global Difference in Host Star Formation}

\author{Young-Lo Kim\altaffilmark{1}, Mathew Smith\altaffilmark{2}, Mark Sullivan\altaffilmark{2}, and Young-Wook Lee\altaffilmark{1}} 

\altaffiltext{1}{Center for Galaxy Evolution Research and Department of Astronomy, Yonsei University, Seoul 03722, Korea; ylkim83@yonsei.ac.kr, ywlee2@yonsei.ac.kr}
\altaffiltext{2}{Department of Physics and Astronomy, University of Southampton, Southampton SO17 1BJ, UK}

\begin{abstract}

It is now established that there is a dependence of the luminosity of type Ia supernovae (SNe Ia) on environment: SNe Ia in young, star-forming, metal-poor stellar populations appear fainter after light-curve shape corrections than those in older, passive, metal-rich environments. This is accounted for in cosmological studies using a global property of the SN host galaxy, typically the host galaxy stellar mass. However, recent low-redshift studies suggest that this effect manifests itself most strongly when using the local star-formation rate (SFR) at the SN location, rather than the global SFR or stellar mass of the host galaxy. At high-redshift, such local SFRs are difficult to determine; here, we show that an equivalent \lq local\rq\ correction can be made by restricting the SN Ia sample in globally star-forming host galaxies to a low-mass host galaxy subset ($\le10^{10}$\,\msun). Comparing this sample of SNe Ia (in locally star-forming environments) to those in locally passive host galaxies, we find that SNe Ia in locally star-forming environments are $0.081\pm0.018$\,mag\ fainter (4.5$\sigma$), consistent with the result reported by \citet{Rigault2015}, but our conclusion is based on a sample $\sim5$ times larger over a wider redshift range. This is a larger difference than when splitting the SN Ia sample based on global host galaxy SFR or host galaxy stellar mass. This method can be used in ongoing and future high-redshift SN surveys, where local SN Ia environments are difficult to determine.

\end{abstract}


\section{Introduction}
\label{sec:intro}

The most direct evidence for the accelerating universe is provided by distances inferred from the measurement of type Ia supernovae \citep[SNe Ia;][]{Riess1998, Perlmutter1999}. The fundamental idea behind the use of SNe Ia is that their luminosities can be empirically standardized \citep{Phillips1993,Tripp1998}, and that standardization does not evolve with redshift or SN environment. Although this assumption was initially supported by small samples of SNe Ia and their host galaxies, which showed no clear luminosity dependence on host galaxy morphology \citep{Riess1998, Schmidt1998, Perlmutter1999, Sullivan2003}, more recent studies with larger numbers of SNe Ia have revealed subtle trends between the SN luminosity after empirical light-curve shape and color/extinction corrections, and host galaxy stellar mass, star formation rate (SFR) and specific SFR (sSFR; the SFR per unit stellar mass), and gas-phase metallicity \citep[e.g.,][]{Kelly2010,  Lampeitl2010, Sullivan2010, D'Andrea2011, Childress2013, Johansson2013, Pan2014}.

In particular, the dependency of the Hubble residual on the host galaxy stellar mass (\mstellar) is well-established: in the recent Joint Lightcurve Analysis (JLA) compilation of 740 SNe Ia \citep{Betoule2014}, SNe Ia in galaxies with $\mstellar\leq10^{10}$\,\msun\ were shown to be $0.061\pm0.012$\,mag fainter than SNe Ia in galaxies with $\mstellar>10^{10}$\msun, after light curve shape and color corrections. The effect presumably arises because of differing properties of the progenitor stellar populations; the leading candidates are the progenitor age and the progenitor metallicity \citep{Timmes2003,Kasen2009}, both correlate with host galaxy stellar mass \citep[e.g.,][]{Tremonti2004,Gallazzi2005, Kang2016}.

These studies were based on measurements of the global properties of the SN Ia host galaxies, with the implicit assumption that the stellar population from which the SN progenitor originated shared these global properties. Although this assumption may be statistically true for large samples of objects, local environmental measurements at the SN location are likely to be more directly linked to the SN progenitor stellar populations. Such local environmental measurements of stellar age or metallicity are, however, difficult to make, and usually require dedicated spectroscopic programmes. One such example, \citet[hereafter R13]{Rigault2013}, used 82 spectroscopic measurements of the SFR at the SN position in the redshift range $0.03<z<0.08$, and found that SNe Ia in locally star-forming environments are $0.094\pm0.031$\,mag fainter than those in locally passive environments. This result was reconfirmed by \citet[hereafter R15]{Rigault2015} using an independent nearby sample of SNe Ia, where the local SFR was estimated instead from \textit{GALEX} far-UV data; in total they estimate a magnitude offset of $0.094\pm0.025$\,mag (see also \citealp{Roman2017} for a similar result from local $U-V$ color). However, \citet[hereafter J15]{Jones2015} reached a different conclusion from the sample based on the same \textit{GALEX} photometry, but including a very nearby sample ($0.01<z<0.023$) and using an updated version of the Spectral Adaptive Lightcurve Template 2 \citep[SALT2;][]{Guy2007, Guy2010}.

It is therefore important to be able to select sample of SNe Ia with known local galaxy properties, perhaps only based on global galaxy measurements, to clarify the effects of local environments on the SN Ia luminosities. The purpose of this paper is to  show that such samples can be efficiently selected when the globally star-forming sample is restricted to relatively low mass hosts ($\le$ 10$^{10}$M$_{\odot}$), and that the environmental dependency of SN Ia luminosities is evident in this much larger sample. Only multi-band photometric data are used in this method, for which data for $\sim$1,000 hosts are available in the literature from low to high redshift ($z<1.1$).

\section{Data}
\label{sec:data}

The SNe Ia data used in this paper are drawn from the YOnsei Nearby Supernova Evolution Investigation (YONSEI) SN Catalog \citep{Kim2015, Kang2016}. This catalog is a superset of all SN Ia surveys adopted in the SNANA package \citep{Kessler2009}, containing 1059 SNe Ia over the redshift range $0.01<z<1.4$, including: low-redshift SN surveys \citep[hereafter \lq Low-$z$\rq]{Hamuy1996, Riess1999, Jha2006, Hicken2009, Hicken2012, Contreras2010, Stritzinger2011}, the SDSS-II SN survey \citep[hereafter SDSS]{Sako2014}, the ESSENCE survey \citep{Miknaitis2007}, the first three years of the Canada-France-Hawaii Telescope Supernova Legacy Survey \citep[hereafter SNLS]{Guy2010}, and the \textit{Hubble Space Telescope} sample of \citet{Riess2007}. For the light-curve analysis, we re-fit all the original published light curves with the most up-to-date version of SALT2 (version 2.4 presented in \citealt{Betoule2014}), as implemented in the SNANA package. The Malmquist bias corrections are then applied to the sample. Because most of the Low-$z$, SDSS, SNLS, and $HST$ samples in the JLA catalog are very similar to those in our catalog, we take the correction terms calculated by \citet{Betoule2014}. For the ESSENCE sample, we adopt the correction values provided by \citet{Wood-Vasey2007}. We then interpolate the bias correction value for each SN at given redshift, and this value is subtracted from all rest-frame peak apparent magnitude in $B$-band ($m_B$). The YONSEI SN Catalog provides a rest-frame peak apparent magnitude in $B$-band ($m_B$), a light-curve shape parameter ($x_1$), and a color parameter ($c$) for each SN. In order to select only normal SNe Ia\footnote{We discard peculiar, sub-, and over-luminous SNe, such as 1991bg-like and 2002cx-like, following the \citet{Betoule2014} scheme.} (the \lq YONSEI Cosmology sample\rq), we apply various cuts to the sample based on the light-curve shape and color values ($-3<x_{1}<3$ and $-0.3<c<0.3$), similar to those adopted in \citet{Betoule2014}. Of the 1059 SNe, 941 pass this requirement. Finally, for our analyses of Hubble residuals and host properties, we restrict the SNLS sample to $z\le0.85$, as in \citet{Sullivan2010}, where the SNLS SNe have the highest signal-to-noise, and the Malmquist corrections are smallest. Full details of our procedures will be published in a companion paper (Y.-L. Kim et al., in preparation).

For the purposes of this paper, which examines SNe Ia in the context of their host galaxies, we also require host galaxy information, specifically stellar masses and global sSFRs. For consistency across the samples, we use the P\'{E}GASE.2 spectral synthesis code \citep{Fioc1997, LeBorgne2004} as described in detail in \citet{Sullivan2006,Sullivan2010}. Briefly, we use a set of 14 exponentially declining star formation histories (SFHs) with SFR $\propto\exp ^{-t/\tau}$, where $t$ represents time and $\tau$ is the e-folding time. Each SFH has 100 time steps, and we use foreground dust screens ranging from $E(B-V)=0$ to 0.30\,mag in steps of 0.05. We fit the host galaxy data from \citet{Sullivan2010} (SNLS) and \citet{Smith2012} and \citet{Sako2014} (SDSS) using the above framework to ensure a consistent approach. 213 hosts for SNLS and 355 for SDSS are matched with the YONSEI Cosmology sample. For the Low-$z$ sample, 89 host information are taken from \citet{Neill2009} which uses the same P\'{E}GASE.2 approach. In total, 657 SNe Ia and their host galaxies are collected for our analysis. The sample sizes and the data used in our analysis are listed in Tables~\ref{tab:number} and~\ref{tab:snproperties}, respectively.

Local host properties for 110 Low-$z$ samples are separately listed in Table~\ref{tab:localproperties}. The local SFR surface density, global host class (e.g., star-forming or passive), and the probability that a SN has a locally passive environment (\probpassive) are taken from \citetalias{Jones2015} and \citetalias{Rigault2015}\footnote{\citetalias{Jones2015} showed a median offset in $\probpassive$ between \citetalias{Jones2015} and \citetalias{Rigault2015} of $\sim3\%$. They showed that this has little impact on the final results.}. Host stellar masses are collected from \citet{Neill2009} and \citet{Kelly2010}.

For the SALT2 model, the distance modulus for each SN is formed as
\begin{equation}
\mu_\mathrm{SN}=m_B-M_B+\alpha\times x_1 -\beta\times c
\end{equation}
where $\alpha$, $\beta$ and $M_B$ are nuisance parameters in the distance modulus estimate. Cosmological fits then minimise
\begin{equation}
\chi^2=\sum_\mathrm{SNe}\frac{\mu_\mathrm{SN}-\mu_\mathrm{model}(z;\Omega_M)}{\sigma^2_\mathrm{stat}+\sigma^2_\mathrm{int}}
\end{equation}
where $\mu_\mathrm{model}(z;\Omega_M)$ is the predicted distance modulus in the $\Lambda$CDM cosmology we assume throughout this paper, $\sigma_\mathrm{stat}$ is the statistical uncertainty for each SN, and $\sigma_\mathrm{int}$ is the so-called intrinsic dispersion added to the SN uncertainties to ensure a reduced $\chi^2$ ($\chi^2_\mathrm{red}$; the $\chi^2$ per degree of freedom) of 1. When we examine the systematic variation in SN Ia luminosity with the host galaxy properties, we set $\sigma_\mathrm{int}=0$ \citep[e.g., see][]{D'Andrea2011, Gupta2011, Pan2014}. We refer to the quantity $\mu_\mathrm{SN}-\mu_\mathrm{model}(z;\Omega_M)$ as the \lq Hubble residual\rq. We use the JLA likelihood code \citep{Betoule2014} to estimate baseline Hubble residuals for the SNe Ia. The best-fit cosmological parameters obtained from the YONSEI Cosmology sample were $\Omega_{M}=0.30$, $\alpha=0.15$, $\beta=3.69$, and $M_{B}=-19.06$ with $\sigma_\mathrm{int}=0$. In the calculation of the weighted-mean of Hubble residuals described below, the error of the weighted-mean is corrected to ensure a $\chi^2_\mathrm{red} = 1$. We have also applied Chauvenet's criterion \citep{Taylor1997} to reject outliers during this procedure, removing 9 objects from our sample. For our analysis, this criterion corresponds to $3.1\sigma$ on average.

For the Low-$z$ sample, the effect of peculiar velocities of the SN host galaxies relative to the Hubble flow may introduce a bias in the determination of the cosmological parameters. Our Low-$z$ sample uses the redshifts corrected for bulk flows \citep[following][]{Conley2011}. Further, \citet{Dhawan2017} and \citet{Zhang2017} showed that this difference has only a negligible effect in measuring the Hubble constant ($\sim0.4\%$). Consequently, we estimate the impact of peculiar velocities on our study -- which is not focused at low-redshift -- is likely to be very small.

\section{Selecting SNe Ia without a local--global difference in star formation}
\label{sec:empiricalmethod}

\begin{figure}
\centering
\includegraphics[angle=0, scale=0.8]{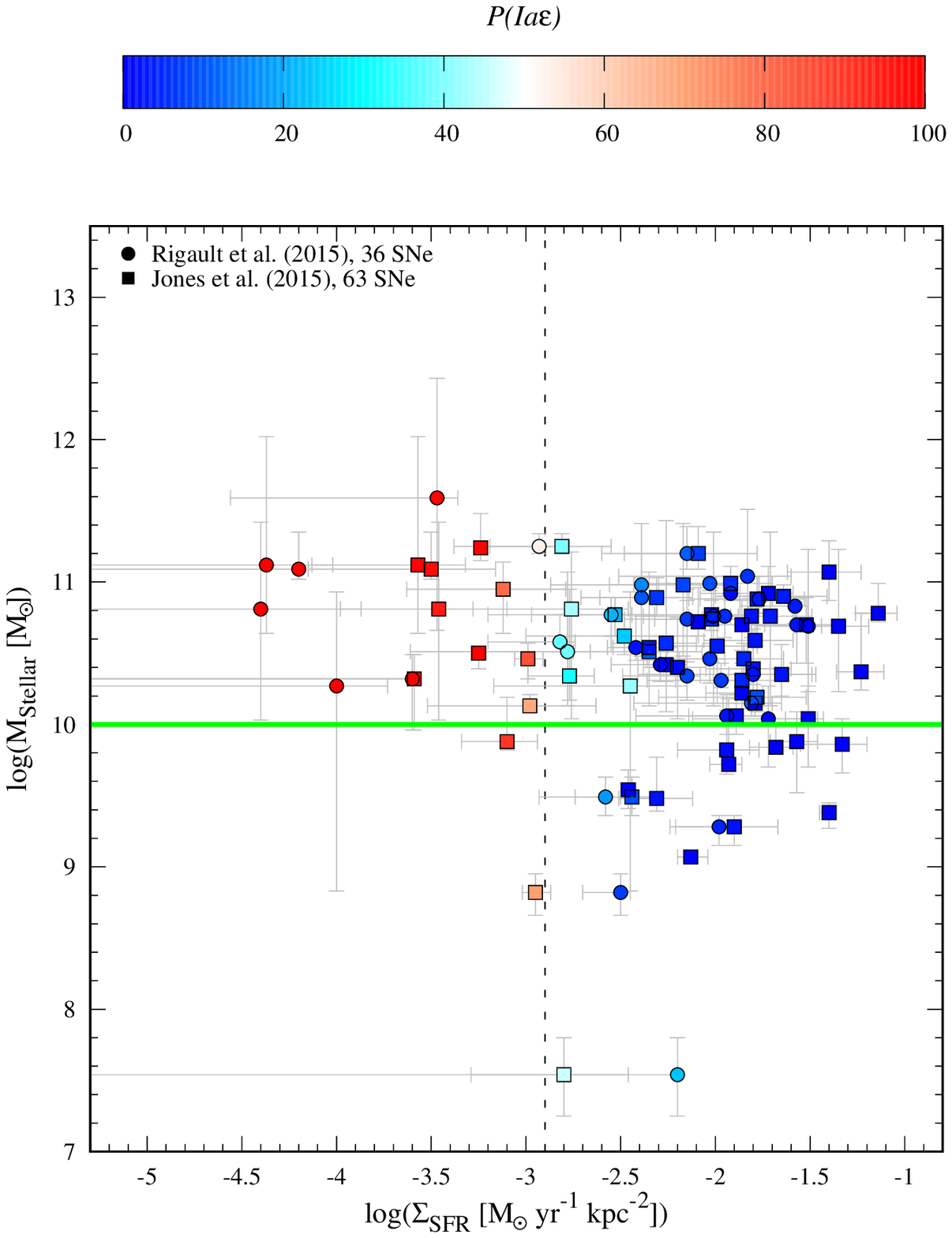}
\caption{Our empirical method for globally star-forming host galaxies to select a SN sample without the local--global difference in star formation. The host galaxy local SFR surface density, $\log(\Sigma_\mathrm{SFR})$, is plotted as a function of host galaxy stellar mass $\log(\mstellar)$, together with the probability of a locally passive environment (\probpassive). Host stellar masses are taken from \citet{Neill2009} and \citet{Kelly2010}. The host local properties are drawn from \citetalias{Rigault2015} (filled circles) and \citetalias{Jones2015} (filled squares), with 32 SNe Ia in common. A clear distinction can be observed when we divide this figure into two regimes at $\log(\mstellar)=10$ (green horizontal line). For the high-mass hosts $\log(\mstellar)>10$, SNe Ia can arise either from locally passive (redder points) or locally star-forming environments (bluer points), while almost all of them occur only in locally star-forming environments for the low-mass hosts $\log(\mstellar)\le10$. Therefore, for the case of globally star-forming hosts, a sample without the local--global difference in star formation can be drawn from the low-mass hosts. SNe Ia are colored by \probpassive. The vertical dashed line shows the local star-formation surface density threshold, $\log(\Sigma_\mathrm{SFR})=-2.9$\,dex, taken from \citetalias{Rigault2015}.\label{fig:empiricalmethod}}
\end{figure}

\citetalias{Rigault2013} showed that SNe Ia occurring in globally passive host galaxies only occur in locally passive environments, but SNe Ia in globally star-forming host galaxies can occur in both locally star-forming and locally passive environments. Cleanly separating the latter class (we shall refer this as the local--global difference in star formation) into locally passive and locally star-forming subsets will clearly be of great benefit. For the globally star-forming host galaxies, we show in Figure~\ref{fig:empiricalmethod} the local SFR surface density ($\log(\Sigma_\mathrm{SFR})$) as a function of the stellar mass (\mstellar) for the Low-$z$ sample listed in Table~\ref{tab:localproperties}. This figure shows that locally star-forming environments are found across a wide range of \mstellar, while the locally passive environments are mostly found in the relatively massive galaxies. Dividing this figure into two regimes at $\log(\mstellar)=10$ gives a clear distinction between these two populations: in high-mass hosts ($\log\mstellar>10$), SNe Ia can arise either from locally passive and locally star-forming environments; in low-mass hosts ($\log\mstellar\le10$), SNe Ia arise only in locally star-forming environments. Thus, for globally star-forming host galaxies, a clean sample of SNe Ia in locally star-forming environments can be formed by selecting only host galaxies with $\log\mstellar\le10$.

With the assumption that this simple empirical criterion can be used for all globally star-forming hosts (with $\log(\mathrm{sSFR})>-10.4$ for our sample), we can apply this to all our globally star-forming hosts to select a sample \textit{without} the local-global difference in star formation (i.e., in the locally star-forming environments). For the SNe Ia in globally passive hosts (with $\log(\mathrm{sSFR})\le-10.4$), we assume all are also in locally passive environments \citepalias[as demonstrated by][]{Rigault2013}. This gives a final sample of 368 SNe Ia (out of 649 in Table~\ref{tab:number}) without the local--global difference in star formation, among which 194 SNe Ia are in locally passive environments ($N_\mathrm{locally-passive}=194$), and 174 SNe Ia are in locally star-forming environments ($N_\mathrm{locally-SF}=174$).

\section{Results}
\label{sec:results}

\begin{figure}
\centering
\includegraphics[angle=0,scale=0.7]{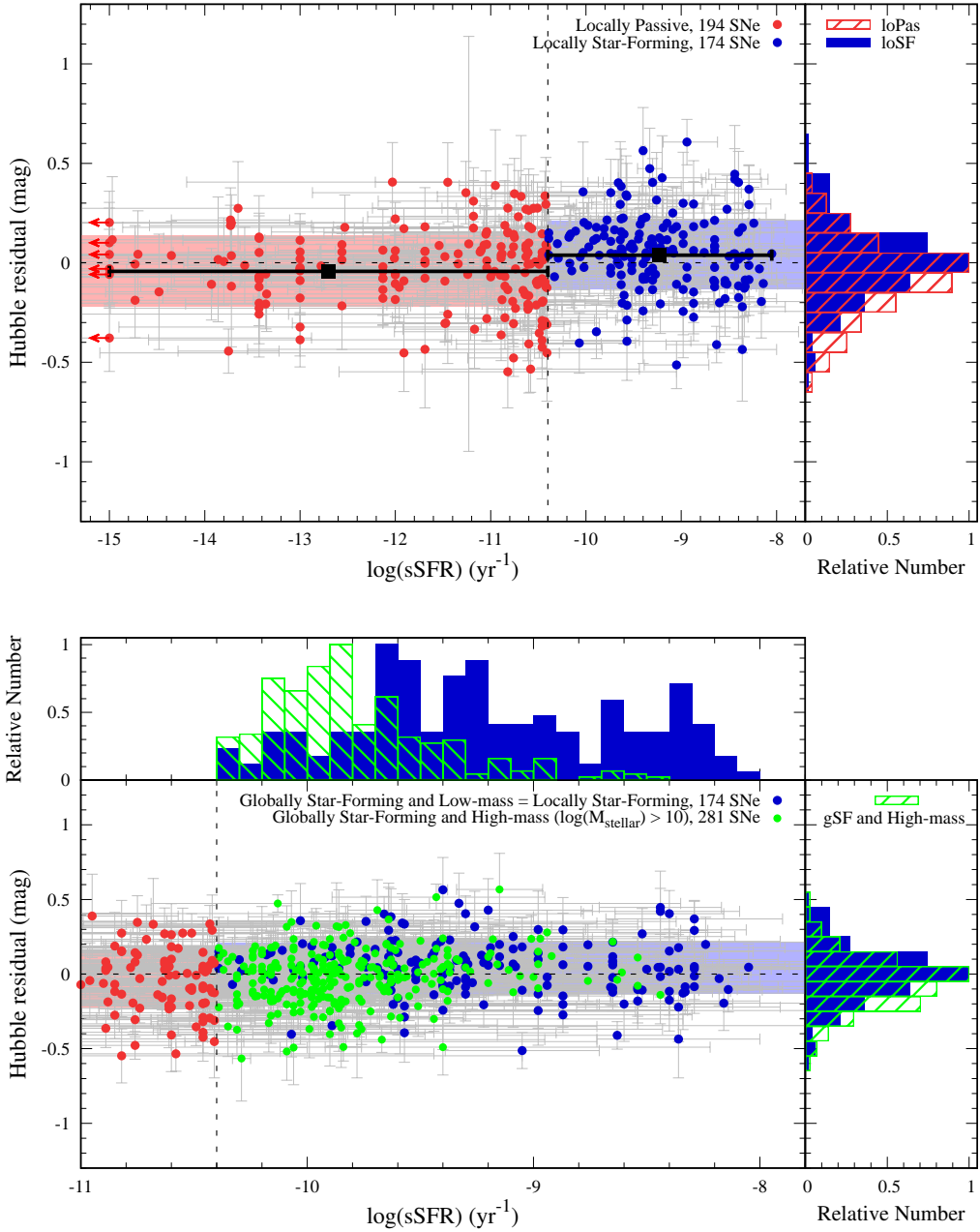}
\caption{
\textit {Upper panel}: Environmental dependence of SN Ia luminosity from a sample without the local--global difference in star formation. The luminosity difference between SNe Ia in locally star-forming environments (blue circles) and those in locally passive environments (red circles) is $0.081\pm0.018$\,mag ($4.5\sigma$). This difference is consistent with the combined result of \citetalias{Rigault2015} ($0.094\pm0.025$\,mag), but is based on a sample 5 times larger and thus statistically more significant. Note that the r.m.s. scatter of SNe Ia in locally star-forming environments (blue shaded area) is $\sim5\%$ smaller than those in locally passive environments (red shaded area). The black squares represent the weighted-mean of Hubble residuals in bins of sSFR. The vertical dotted line indicates the limit distinguishing between passive and star-forming galaxies for our sample. \textit{Lower panel}: The distribution of SNe Ia in globally star-forming and high-mass ($\log(\mstellar)>10$) hosts (green circles), which have a sample \textit{with} the local--global difference in star formation. They have more negative Hubble residuals (right panel) and their hosts show less SFR than globally star-forming and low-mass (i.e., locally star-forming) hosts (upper panel).\label{fig:results}
}
\end{figure}

The upper panel of Figure~\ref{fig:results} shows the dependence of SN Ia luminosity on host galaxy properties from the sample without the local--global difference in star formation, as described in Section~\ref{sec:empiricalmethod}. We find that SNe Ia in locally star-forming environments are fainter than those in locally passive environments: the difference in the weighted-mean of the Hubble residuals is $0.081\pm0.018$\,mag (see Table~\ref{tab:hostresults}). These values are comparable to the combined result of \citetalias{Rigault2015}: $0.094\pm0.025$\,mag ($3.5\sigma$) for SALT2. Our result, however, is statistically more significant ($4.5\sigma$), as the present analysis is based on a $\sim5$ times larger sample that covers a wider redshift range ($0.01<z\le0.85$). This is an independent confirmation of the environmental dependency of SN Ia luminosity from the sample without the local-global difference in star formation -- but based only on multi-band photometry.

We contrast this result to that based on our full YONSEI Cosmology sample. The lower panel of Figure~\ref{fig:results} also shows this cosmology sample as a function of sSFR, but including all SNe Ia in star-forming galaxies, regardless of their stellar mass. The sample of SNe Ia in globally star-forming galaxies with $\log\mstellar>10$ have more negative Hubble residuals (i.e., are brighter, $-0.009\pm0.010$) than those SNe Ia in globally star-forming galaxies with $\log\mstellar\le10$ ($0.038\pm0.013$). The difference in Hubble residual between globally star-forming and globally passive galaxies is reduced to $0.049\pm0.015$\,mag (see Table~\ref{tab:hostresults}). We also compare the results when splitting the sample according to \mstellar\ (see Table~\ref{tab:hostresults}). For such a sample, the difference in Hubble residual is not as large as the result from a sample without the local--global difference. 

We check the probability of observing our main result by chance using a Monte Carlo permutation test. We randomly draw $N_\mathrm{locally-SF}$ SNe Ia from our full star-forming sample (without replacement) and calculate the difference in Hubble residuals between this randomly selected star-forming sample and the globally passive/locally passive hosts. In 100,000 realizations, $\simeq0.4\%$ of the samples have a larger value for the difference in Hubble residuals than our main result.

As discussed by \citetalias{Jones2015} and \citet{Rigault2015Conf}, the redshift cut may affect their results, which show an apparent discrepancy (see Section~\ref{sec:intro}). In order to investigate this effect, we have selected a sample overlapping in redshift with \citetalias{Jones2015} ($0.010 < z < 0.1$) and \citetalias{Rigault2015} ($0.023 < z < 0.1$). In the case of a sample overlapping with \citetalias{Jones2015} (67 SNe), we also observe no significant environmental dependency in the SN luminosity: the luminosity differences are $0.011\pm0.042$\,mag. For the sample overlapping in redshift with \citetalias{Rigault2015} (43 SNe), the luminosity differences are  $0.080\pm0.046$\,mag (1.7$\sigma$). These results are consistent, if at slightly lower significance than those of \citetalias{Rigault2015}, who found differences at 2.5$\sigma$. For our full redshift sample only $\sim6\%$ of SNe have $z < 0.023$, so the redshift cut has only negligible effect on our main result.

\section{Implication for Cosmology}
\label{sec:cosmology}

The environmental dependency of SN Ia luminosity obtained in this paper is consistent with a link between SN Ia progenitor age and SN Ia standardised luminosity: those in passive environments, brighter after light-curve shape correction, and those in star-forming environments, fainter after light-curve shape correction. As the cosmic star-formation history evolves sharply with redshift, the mix of these SNe Ia is also likely to change, with the fraction of SNe in locally passive environments most likely decreasing with increasing redshift \citep[e.g.][]{Sullivan2006}. In order to predict the impact on cosmology, we have employed a simple model for the fraction of SNe Ia located in locally passive environments as a function of redshift, $\psi(z)$, from \citetalias{Rigault2013} (their Eq.(5)), defined as
\begin{equation}
\psi(z)= (K \times 10^{0.95z} + 1)^{-1},
\end{equation}
with $K=0.90\pm0.15$ from the normalization of $\psi(z=0.05) = 50\pm5\%$, based on their observational data. The difference in Hubble residual between SNe in locally passive and those in locally star-forming environments as a function of redshift can then be written as
\begin{equation} \label{eq:delHR}
\Delta \mathrm{HR_{SFR}}(z) = A \times \psi(z).
\end{equation}
For the calibration of $\Delta\mathrm{HR}_{\mathrm{SFR}}(z)$, we split our data into three redshift bins with equal numbers of SNe ($z \le 0.180$, $0.180 < z < 0.337$, and $0.337 < z$, $\sim123$ SNe in each bin), from which we measure $A$ to be $-0.204\pm0.035$. Figure~\ref{fig:delHRvsz} shows the evolution of environmental dependence of SN luminosity from these three redshift bins, which is compared with the simple evolution model from Eq.~(\ref{eq:delHR}). The model is consistent with our observed data, similar to the study of \citetalias{Rigault2013}, but that study is based on the mass-step evolution model ($\Delta M_{B, \mathrm{mass}}^\mathrm{corr}(z)$, see their Figure 11). Clearly, further SN Ia data are required to test the model in detail. \citetalias{Rigault2013} pointed out that ignoring any observed redshift evolution of the HR difference could shift the dark energy equation-of-state by $\Delta w \sim -0.06$; our result would suggest a similar shift in $w$.

The sample without the local--global difference in star formation can also give the more robust results when estimating cosmological parameters, in terms of the r.m.s. scatter of the Hubble residuals (Figure~\ref{fig:results} and Table~\ref{tab:hostresults}) and the intrinsic scatter (Table~\ref{tab:cosmology}). Interestingly, SNe Ia in locally star-forming environments have a $2\%$ smaller r.m.s. scatter, and also require a  $5\%$ smaller intrinsic scatter than those in the full local sample. This may indicate that this sample is made up of the most homogeneous sample in terms of progenitor ages, and therefore less affected by a possible luminosity evolution of SNe Ia.

\section{Discussion}
\label{sec:discussion}

We have shown that a sample without the local-global difference in star formation can be efficiently selected, when the globally star-forming sample is restricted to the relatively low-mass hosts ($\le$ 10$^{10}$M$_{\odot}$). By employing this technique, we find that SNe Ia in locally star-forming environments are $0.081\pm0.018$\,mag  fainter than those in locally passive environments, after light-curve shape and color corrections. When only the lowest redshift bin ($z\le0.180$) is considered, this luminosity difference increases slightly to $0.091\pm0.031$\,mag. This is consistent with the results suggested by \citetalias{Rigault2013} and \citetalias{Rigault2015}. Our results are, however, statistically more significant (4.5$\sigma$) than previous results, because our sample is $\sim$5 times larger.

As noted above, a clear distinction in the local--global difference in star formation is observed when the sample is divided at $\log(\mstellar)=10$. The well-established mass-step in SN Ia luminosity also occurs near this host mass \citep{Kelly2010, Lampeitl2010, Sullivan2010, Gupta2011, Childress2013, Betoule2014, Pan2014}. Numerous studies pointed out the uniqueness of the mass scale of 10$^{10}$M$_{\odot}$. For example, \citet{Cappellari2013} and \citet{Bernardi2016} showed that $\mstellar\sim3\times10^{10}$\,\msun\ is related to a transition in the assembly histories of galaxies for both early- and late-type galaxies. Furthermore, \citet{Kauffmann2003}, \citet{Balcells2007}, and \citet{Hopkins2009} showed a transition of galaxy morphology, such that galaxy morphology is changed from late- to early-type or from a disk to a bulge-dominated system, occurs near this mass scale. These results suggest that the origin of luminosity difference between SNe Ia in star-forming (low-mass) and those in passive (high-mass) hosts may be related to these transitions, because the average mass, metallicity, and population age of hosts change as well. In particular, in the case of globally star-forming low-mass ($\log\mstellar\le10$) galaxies, the star formation history is characterized by recent starbursts with little contribution from older stars. By contrast, in globally star-forming high-mass galaxies, the star formation history appears more extended, with more contributions from older stars \citep[e.g.,][]{Kauffmann2003, Salim2007}. Therefore, SNe Ia in globally star-forming low-mass hosts are more likely originating from young progenitors \citep[see also figure 3 of][]{Childress2014}.

\begin{figure}
\centering
\includegraphics[angle=-90,scale=0.85]{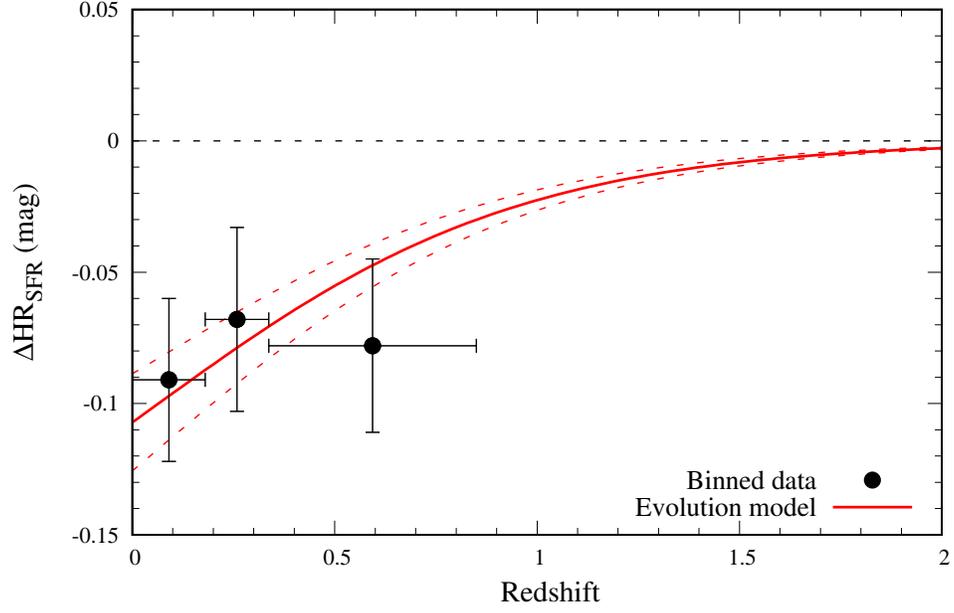}
\caption{The observed and predicted evolution of the difference in Hubble residual between SNe Ia in locally passive and SNe Ia in locally star-forming environments ($\Delta \mathrm{HR}_\mathrm{SFR}$). The filled circles are the observed $\Delta \mathrm{HR}_\mathrm{SFR}$ in three redshift bins from this study. The red solid line shows the simple evolution model from Eq. (\ref{eq:delHR}). This model is consistent with our observed data, but further SN Ia data are required to test the model in detail. The red dotted lines show $\pm1\sigma$ ranges of the evolution model. \label{fig:delHRvsz}
}
\end{figure}

Our result on the luminosity difference between SNe Ia in locally passive and those in locally star-forming environments is qualitatively consistent with the well-established mass-step in SN Ia luminosity. Since the host mass and SFR cannot directly affect SN luminosity, many studies pointed out that this is most likely due to the population properties of a host galaxy, such as age and metallicity \citep{Johansson2013, Childress2014, Pan2014, Graur2015, Kang2016}. Specifically, \citet{Kang2016} found that stellar population age is mainly responsible for the relation between host mass and Hubble residual. This would imply that the properties of SN can vary with a mean population age of a host galaxy. As this quantity is known to evolve with redshift, the properties of a progenitor would also change with redshift. This in turn may affect the details of SN explosion mechanism, and therefore would lead to a possible luminosity evolution of SNe Ia, as highlighted in Figure~\ref{fig:delHRvsz}.  Since the luminosity evolution can cause a potential bias in the estimation of cosmological parameters, it deserves a careful consideration when using SN Ia for cosmological analyses. 

Finally, we note that, even though $\sim43\%$ of the full local sample is not used after performing the analysis suggested in this paper, SNe Ia in locally star-forming environments could give more robust results: a $\sim2\%$ smaller r.m.s. scatter of the Hubble residual and a $\sim5\%$ smaller intrinsic scatter than when using the full local sample (see Section~\ref{sec:cosmology}). As has also been suggested by \citet{Rigault2013}, \citet{Childress2014}, and \citet{Kelly2015}, this homogeneous sample can also lead to the improved application of SNe Ia as cosmological distance indicators. Therefore, future SN Ia cosmology surveys should consider the measurements of the local environments of host galaxies. The local properties of hosts, however, cannot be directly determined from spectroscopy, even in the era of 30-m class telescopes. In this respect, the method presented in this paper, which requires only global multi-band host galaxy photometry, could be adopted in forthcoming high-redshift SN surveys.

\acknowledgments

We thank the referee for his/her careful reading of the manuscript and many helpful comments. Y.-L. K. thanks the Department of Physics and Astronomy at the University of Southampton for their hospitality during the visits, and Mario Pasquato for helpful suggestions on the MCMC test. M.S. acknowledges support from EU/FP7-ERC grant No. [615929]. Support for this work was provided by the National Research Foundation of Korea to the Center for Galaxy Evolution Research through the grant programs No. 2017R1A5A1070354 and 2017R1A2B3002919.

\clearpage

\clearpage

\clearpage

\clearpage

\begin{deluxetable}{ccccc}
\tabletypesize{\scriptsize}
\tablewidth{0pt}
\tablecaption{\label{tab:number}The sample sizes for each group of host galaxies.}
\tablehead{
\colhead{YONSEI Cosmology} & \colhead{} & \multicolumn{3}{c}{Host groups} \\
\cline{3-5}
\colhead{} & \colhead{} & \colhead{Mass} & \colhead{Global sSFR} & \colhead{Local group from this work}
}
\startdata
941 & & 648 (657) & 649 (657) & 368 (373) \\
\enddata
\tablecomments{The number in parenthesis is the value before applying Chauvenet's criterion (see Section~\ref{sec:data}).}
\end{deluxetable}

\clearpage

\floattable
\begin{deluxetable}{lcccccccccccc}
\tabletypesize{\scriptsize}
\tablewidth{0pt}
\tablecaption{\label{tab:snproperties} YONSEI Supernova Catalog and host \lq global\rq\ properties}
\tablehead{
\colhead{} & \colhead{} & \colhead{} & \multicolumn{2}{c}{SALT2} & \colhead{} & \multicolumn{3}{c}{Host Mass} & \colhead{} & \multicolumn{3}{c}{Global sSFR} \\
\cline{4-5} \cline{7-9} \cline{11-13}
\colhead{Name} & \colhead{Survey} & \colhead{z$_{CMB}$} & \colhead{HR} & \colhead{$\sigma$$_{HR}$} & \colhead{} & \colhead{log(M$_{stellar}$)} & \colhead{$-$$\delta$} & \colhead{$+$$\delta$} & \colhead{} & \colhead{log(sSFR)} & \colhead{$-$$\delta$} & \colhead{$+$$\delta$} \\
\colhead{} & \colhead{} & \colhead{} & \colhead{(mag)} & \colhead{} & \colhead{} & \colhead{(M$_{\odot}$)} & \colhead{} & \colhead{} & \colhead{} & \colhead{(yr$^{-1}$)} & \colhead{} & \colhead{}
}
\startdata
1991ag	&	LOWZ	&	0.014 	&	-0.088 	&	0.122 	&	&	9.07 	&	0.03 	&	0.03 	&	&	-8.66 	&	0.06 	&	0.08 	\\
1992P	&	LOWZ	&	0.026 	&	0.134 	&	0.161 	&	&	10.34 	&	0.10 	&	0.14 	&	&	-9.98 	&	0.88 	&	0.63 	\\
1992ag	&	LOWZ	&	0.026 	&	-0.714 	&	0.156 	&	&	10.02 	&	0.10 	&	1.04 	&	&	-9.88 	&	1.23 	&	0.93 	\\
1992bc	&	LOWZ	&	0.020 	&	0.063 	&	0.101 	&	&	9.72 	&	0.53 	&	0.70 	&	&	-9.62 	&	0.94 	&	1.12 	\\
1992bl	&	LOWZ	&	0.043 	&	-0.069 	&	0.168 	&	&	11.81 	&	0.65 	&	0.46 	&	&	-10.85 	&	0.71 	&	1.58 	\\
722	&	SDSS3	&	0.085 	&	0.055 	&	0.115 	&	&	10.93 	&	0.01 	&	0.01 	&	&	-12.56 	&	0.70 	&	0.70 	\\
739	&	SDSS3	&	0.106 	&	-0.117 	&	0.151 	&	&	11.17 	&	0.03 	&	0.03 	&	&	-10.85 	&	0.17 	&	0.17 	\\
744	&	SDSS3	&	0.127 	&	-0.061 	&	0.133 	&	&	10.62 	&	0.11 	&	0.11 	&	&	-10.24 	&	0.22 	&	0.22 	\\
762	&	SDSS3	&	0.190 	&	0.175 	&	0.120 	&	&	11.23 	&	0.01 	&	0.01 	&	&	-10.39 	&	0.40 	&	0.40 	\\
774	&	SDSS3	&	0.092 	&	-0.007 	&	0.101 	&	&	10.81 	&	0.02 	&	0.02 	&	&	-10.31 	&	0.10 	&	0.10 	\\
03D1ar	&	SNLS3	&	0.408 	&	0.066 	&	0.135 	&	&	10.46 	&	0.18 	&	0.18 	&	&	-8.98 	&	0.39 	&	0.39 	\\
03D1au	&	SNLS3	&	0.504 	&	-0.061 	&	0.115 	&	&	9.75 	&	0.04 	&	0.04 	&	&	-9.63 	&	0.13 	&	0.13 	\\
03D1aw	&	SNLS3	&	0.582 	&	0.177 	&	0.141 	&	&	9.57 	&	0.17 	&	0.17 	&	&	-9.19 	&	0.26 	&	0.26 	\\
03D1ax	&	SNLS3	&	0.496 	&	-0.016 	&	0.110 	&	&	11.70 	&	0.08 	&	0.08 	&	&	-12.78 	&	0.68 	&	0.68 	\\
03D1bp	&	SNLS3	&	0.347 	&	0.133 	&	0.103 	&	&	11.03 	&	0.02 	&	0.02 	&	&	-10.20 	&	0.06 	&	0.06 	\\
\enddata
\tablecomments{Table~\ref{tab:snproperties} is published in its entirety in the machine-readable format. A portion is shown here for guidance regarding its form and content.}
\end{deluxetable}

\clearpage

\begin{deluxetable}{lcccccccccccccc}
\tabletypesize{\scriptsize}
\tablewidth{0pt}
\tablecaption{\label{tab:localproperties} Host stellar mass and \lq local\rq\ properties for the Low-$z$ sample}
\tablehead{
\colhead{} & \colhead{} & \multicolumn{3}{c}{Host Mass} & \colhead{Global} & \multicolumn{4}{c}{\citetalias{Rigault2015} Local SFR} & \colhead{} & \multicolumn{4}{c}{\citetalias{Jones2015} Local SFR} \\
\cline{3-5} \cline{7-10} \cline{12-15}
\colhead{Name} & \colhead{z$_{CMB}$} & \colhead{log(M$_{stellar}$)} & \colhead{$-$$\delta$} & \colhead{$+$$\delta$} & \colhead{Host Class} & \colhead{log($\Sigma$$_{SFR}$)} & \colhead{$-$$\delta$} & \colhead{$+$$\delta$} & \colhead{\textit{P(Ia$\epsilon$)}} & \colhead{} & \colhead{log($\Sigma$$_{SFR}$)} & \colhead{$-$$\delta$} & \colhead{$+$$\delta$} & \colhead{\textit{P(Ia$\epsilon$)}} \\
\colhead{} & \colhead{} & \colhead{(M$_{\odot}$)} & \colhead{} & \colhead{} & \colhead{} & \colhead{(M$_{\odot}$ yr$^{-1}$ kpc$^{-2}$)} & \colhead{} & \colhead{} & \colhead{($\%$)} & \colhead{} &  \colhead{(M$_{\odot}$ yr$^{-1}$ kpc$^{-2}$)} & \colhead{} & \colhead{} & \colhead{($\%$)}
}
\startdata
1991U	&	0.033 	&	11.04 	&	0.70 	&	0.47 	&	SF	&	-1.83 	&	0.68 	&	0.23 	&	6 	&	&	\nodata	&	\nodata	&	\nodata	&	\nodata	\\
1991ag	&	0.014 	&	9.07 	&	0.03 	&	0.03 	&	SF	&	\nodata	&	\nodata	&	\nodata	&	\nodata	&	&	-2.13 	&	0.07 	&	0.09 	&	0 	\\
1992bl	&	0.043 	&	11.81 	&	0.65 	&	0.46 	&	Pa	&	-3.48 	&	99.00 	&	0.43 	&	96 	&	&	-3.05 	&	0.46 	&	0.23 	&	80 	\\
1992bo	&	0.018 	&	12.13 	&	1.20 	&	0.14 	&	Pa	&	\nodata	&	\nodata	&	\nodata	&	\nodata	&	&	-4.28 	&	99.00 	&	0.00 	&	100 	\\
1993H	&	0.025 	&	10.51 	&	0.38 	&	0.56 	&	SF	&	-2.78 	&	0.83 	&	0.12 	&	39 	&	&	-2.35 	&	0.40 	&	0.28 	&	11 	\\
1993ac	&	0.049 	&	11.44 	&	0.03 	&	0.04 	&	Pa	&	-4.24 	&	99.00 	&	0.30 	&	98 	&	&	-2.73 	&	0.61 	&	0.32 	&	51 	\\
1993ae	&	0.018 	&	10.35 	&	0.06 	&	1.55 	&	Pa	&	\nodata	&	\nodata	&	\nodata	&	\nodata	&	&	-4.54 	&	99.00 	&	0.00 	&	100 	\\
1994M	&	0.024 	&	11.04 	&	0.11 	&	0.19 	&	Pa	&	-4.30 	&	99.00 	&	0.00 	&	100 	&	&	-3.34 	&	99.00 	&	0.00 	&	100 	\\
1994Q	&	0.029 	&	9.84 	&	0.04 	&	0.16 	&	SF	&	\nodata	&	\nodata	&	\nodata	&	\nodata	&	&	-1.68 	&	0.14 	&	0.09 	&	0 	\\
1994S	&	0.016 	&	10.50 	&	0.11 	&	0.03 	&	SF	&	\nodata	&	\nodata	&	\nodata	&	\nodata	&	&	-3.25 	&	0.04 	&	0.03 	&	100 	\\
1994T	&	0.036 	&	9.60 	&	0.37 	&	1.00 	&	Pa	&	-4.30 	&	99.00 	&	0.00 	&	100 	&	&	-3.47 	&	0.69 	&	0.28 	&	98 	\\
1996bo	&	0.016 	&	10.37 	&	0.13 	&	0.41 	&	SF	&	\nodata	&	\nodata	&	\nodata	&	\nodata	&	&	-1.23 	&	0.13 	&	0.12 	&	0 	\\
1997cn	&	0.017 	&	11.42 	&	0.04 	&	0.04 	&	Pa	&	\nodata	&	\nodata	&	\nodata	&	\nodata	&	&	-3.29 	&	0.17 	&	0.18 	&	99 	\\
1998ab	&	0.028 	&	10.59 	&	0.04 	&	0.23 	&	SF	&	\nodata	&	\nodata	&	\nodata	&	\nodata	&	&	-1.79 	&	0.27 	&	0.20 	&	2 	\\
1998de	&	0.016 	&	11.25 	&	0.38 	&	0.65 	&	Pa	&	\nodata	&	\nodata	&	\nodata	&	\nodata	&	&	-3.41 	&	99.00 	&	0.00 	&	100 	\\
1998dx	&	0.054 	&	11.72 	&	0.55 	&	0.87 	&	Pa	&	-3.70 	&	99.00 	&	0.00 	&	97 	&	&	-2.77 	&	99.00 	&	0.00 	&	93 	\\
1998ec	&	0.020 	&	10.57 	&	0.48 	&	0.86 	&	SF	&	\nodata	&	\nodata	&	\nodata	&	\nodata	&	&	-2.26 	&	0.29 	&	0.16 	&	5 	\\
1999X	&	0.026 	&	10.13 	&	0.03 	&	0.08 	&	SF	&	\nodata	&	\nodata	&	\nodata	&	\nodata	&	&	-2.98 	&	0.54 	&	0.35 	&	68 	\\
1999aa	&	0.015 	&	10.72 	&	0.10 	&	0.24 	&	SF	&	\nodata	&	\nodata	&	\nodata	&	\nodata	&	&	-2.09 	&	0.08 	&	0.06 	&	0 	\\
1999cc	&	0.032 	&	10.99 	&	0.05 	&	0.04 	&	SF	&	-2.03 	&	0.40 	&	0.09 	&	9 	&	&	-1.92 	&	0.34 	&	0.22 	&	5 	\\
1999cp	&	0.010 	&	9.48 	&	0.09 	&	0.29 	&	SF	&	\nodata	&	\nodata	&	\nodata	&	\nodata	&	&	-2.31 	&	0.20 	&	0.19 	&	2 	\\
1999da	&	0.013 	&	10.91 	&	0.12 	&	0.22 	&	Pa	&	\nodata	&	\nodata	&	\nodata	&	\nodata	&	&	-4.02 	&	0.53 	&	0.32 	&	100 	\\
1999dq	&	0.014 	&	10.78 	&	0.06 	&	0.21 	&	SF	&	\nodata	&	\nodata	&	\nodata	&	\nodata	&	&	-1.14 	&	0.12 	&	0.10 	&	0 	\\
2000ca	&	0.025 	&	10.04 	&	0.34 	&	0.23 	&	SF	&	-1.72 	&	0.40 	&	0.26 	&	0 	&	&	-1.51 	&	0.12 	&	0.08 	&	0 	\\
2000dk	&	0.016 	&	11.54 	&	1.37 	&	0.02 	&	Pa	&	\nodata	&	\nodata	&	\nodata	&	\nodata	&	&	-3.75 	&	0.07 	&	0.06 	&	100 	\\
2000fa	&	0.022 	&	9.82 	&	0.17 	&	0.28 	&	SF	&	\nodata	&	\nodata	&	\nodata	&	\nodata	&	&	-1.94 	&	0.26 	&	0.17 	&	1 	\\
2001N	&	0.022 	&	10.77 	&	0.11 	&	0.19 	&	SF	&	\nodata	&	\nodata	&	\nodata	&	\nodata	&	&	-2.02 	&	0.30 	&	0.21 	&	3 	\\
2001ay	&	0.031 	&	10.62 	&	0.13 	&	0.15 	&	SF	&	\nodata	&	\nodata	&	\nodata	&	\nodata	&	&	-2.48 	&	0.52 	&	0.39 	&	24 	\\
2001ba	&	0.031 	&	10.98 	&	0.50 	&	0.43 	&	SF	&	-2.39 	&	0.48 	&	0.09 	&	16 	&	&	-2.17 	&	0.39 	&	0.28 	&	8 	\\
2001en	&	0.016 	&	10.38 	&	0.15 	&	0.15 	&	Pa	&	\nodata	&	\nodata	&	\nodata	&	\nodata	&	&	-2.30 	&	0.30 	&	0.23 	&	8 	\\
2001fe	&	0.014 	&	10.22 	&	0.11 	&	0.12 	&	SF	&	\nodata	&	\nodata	&	\nodata	&	\nodata	&	&	-1.86 	&	0.19 	&	0.11 	&	0 	\\
2001gb	&	0.027 	&	10.39 	&	0.11 	&	0.28 	&	SF	&	\nodata	&	\nodata	&	\nodata	&	\nodata	&	&	-1.80 	&	0.25 	&	0.20 	&	2 	\\
2001ic	&	0.043 	&	11.70 	&	0.06 	&	0.04 	&	Pa	&	-5.80 	&	99.00 	&	0.00 	&	100 	&	&	-2.30 	&	99.00 	&	0.00 	&	77 	\\
2001ie	&	0.031 	&	10.99 	&	0.07 	&	0.21 	&	Pa	&	-3.90 	&	99.00 	&	0.00 	&	100 	&	&	-3.68 	&	0.65 	&	0.35 	&	99 	\\
2002G	&	0.035 	&	10.77 	&	0.15 	&	0.12 	&	SF	&	-2.55 	&	0.65 	&	0.11 	&	24 	&	&	-2.53 	&	0.44 	&	0.36 	&	21 	\\
2002bf	&	0.025 	&	10.62 	&	0.07 	&	0.12 	&	Pa	&	-2.18 	&	0.48 	&	0.10 	&	12 	&	&	-2.23 	&	0.37 	&	0.22 	&	10 	\\
2002de	&	0.028 	&	10.83 	&	0.12 	&	0.03 	&	SF	&	-1.58 	&	0.29 	&	0.07 	&	3 	&	&	\nodata	&	\nodata	&	\nodata	&	\nodata	\\
2002dp	&	0.011 	&	10.40 	&	0.36 	&	0.35 	&	SF	&	\nodata	&	\nodata	&	\nodata	&	\nodata	&	&	-2.20 	&	0.18 	&	0.15 	&	1 	\\
2002ha	&	0.013 	&	11.09 	&	0.13 	&	0.14 	&	Pa	&	\nodata	&	\nodata	&	\nodata	&	\nodata	&	&	-2.53 	&	0.21 	&	0.14 	&	6 	\\
2002he	&	0.025 	&	11.12 	&	0.48 	&	0.90 	&	SF	&	-4.37 	&	99.00 	&	0.35 	&	100 	&	&	-3.57 	&	0.56 	&	0.25 	&	99 	\\
2002hu	&	0.038 	&	10.27 	&	1.44 	&	0.66 	&	SF	&	-4.00 	&	99.00 	&	0.27 	&	100 	&	&	-2.45 	&	0.72 	&	0.30 	&	42 	\\
2002hw	&	0.016 	&	10.38 	&	0.14 	&	0.18 	&	Pa	&	\nodata	&	\nodata	&	\nodata	&	\nodata	&	&	-2.17 	&	0.33 	&	0.23 	&	6 	\\
2002jy	&	0.019 	&	10.46 	&	0.14 	&	0.11 	&	SF	&	\nodata	&	\nodata	&	\nodata	&	\nodata	&	&	-2.99 	&	0.07 	&	0.08 	&	85 	\\
2003U	&	0.028 	&	10.74 	&	0.30 	&	0.13 	&	SF	&	-2.15 	&	0.33 	&	0.08 	&	7 	&	&	-2.02 	&	0.30 	&	0.19 	&	4 	\\
2003W	&	0.021 	&	10.55 	&	0.40 	&	0.25 	&	SF	&	\nodata	&	\nodata	&	\nodata	&	\nodata	&	&	-1.99 	&	0.26 	&	0.21 	&	3 	\\
2003cq	&	0.034 	&	11.20 	&	0.11 	&	0.19 	&	SF	&	-2.15 	&	0.45 	&	0.09 	&	11 	&	&	-2.09 	&	0.39 	&	0.31 	&	7 	\\
2003fa	&	0.039 	&	10.81 	&	0.78 	&	0.61 	&	SF	&	-4.40 	&	99.00 	&	0.53 	&	100 	&	&	-3.46 	&	0.52 	&	0.36 	&	95 	\\
2003hu	&	0.075 	&	10.90 	&	0.04 	&	0.08 	&	SF	&	\nodata	&	\nodata	&	\nodata	&	\nodata	&	&	-1.64 	&	0.36 	&	0.23 	&	3 	\\
2003ic	&	0.054 	&	11.70 	&	0.03 	&	0.27 	&	Pa	&	-3.56 	&	0.18 	&	0.05 	&	100 	&	&	-3.37 	&	0.19 	&	0.08 	&	100 	\\
2004L	&	0.033 	&	10.35 	&	0.15 	&	0.21 	&	SF	&	-1.80 	&	0.61 	&	0.31 	&	2 	&	&	-1.65 	&	0.28 	&	0.18 	&	2 	\\
2004as	&	0.032 	&	9.28 	&	0.13 	&	0.08 	&	SF	&	-1.98 	&	0.26 	&	0.07 	&	4 	&	&	-1.90 	&	0.31 	&	0.23 	&	2 	\\
2005eq	&	0.028 	&	10.58 	&	0.07 	&	0.19 	&	SF	&	-2.82 	&	0.38 	&	0.08 	&	38 	&	&	\nodata	&	\nodata	&	\nodata	&	\nodata	\\
2005hc	&	0.045 	&	10.54 	&	0.09 	&	0.23 	&	SF	&	-2.42 	&	0.15 	&	0.04 	&	2 	&	&	-2.35 	&	0.19 	&	0.15 	&	2 	\\
2005hj	&	0.057 	&	9.49 	&	0.13 	&	0.14 	&	SF	&	-2.58 	&	0.35 	&	0.08 	&	18 	&	&	-2.44 	&	0.30 	&	0.16 	&	9 	\\
2005hk	&	0.012 	&	9.54 	&	0.13 	&	0.14 	&	SF	&	\nodata	&	\nodata	&	\nodata	&	\nodata	&	&	-2.46 	&	0.04 	&	0.04 	&	0 	\\
2005iq	&	0.033 	&	10.34 	&	0.17 	&	0.61 	&	SF	&	-2.15 	&	0.34 	&	0.07 	&	8 	&	&	-2.77 	&	0.22 	&	0.13 	&	32 	\\
2005ir	&	0.075 	&	10.15 	&	0.09 	&	0.05 	&	SF	&	-1.81 	&	0.62 	&	0.11 	&	12 	&	&	-1.79 	&	0.17 	&	0.14 	&	0 	\\
2005kc	&	0.014 	&	10.97 	&	0.13 	&	0.12 	&	Pa	&	\nodata	&	\nodata	&	\nodata	&	\nodata	&	&	-1.76 	&	0.22 	&	0.17 	&	0 	\\
2005ls	&	0.021 	&	9.86 	&	0.20 	&	0.18 	&	SF	&	\nodata	&	\nodata	&	\nodata	&	\nodata	&	&	-1.33 	&	0.22 	&	0.13 	&	0 	\\
2005mc	&	0.026 	&	10.95 	&	0.09 	&	0.11 	&	Pa	&	-2.54 	&	0.31 	&	0.07 	&	14 	&	&	-3.68 	&	0.16 	&	0.11 	&	100 	\\
2005ms	&	0.026 	&	10.32 	&	0.36 	&	0.17 	&	SF	&	-3.60 	&	99.00 	&	0.00 	&	98 	&	&	-3.59 	&	99.00 	&	0.00 	&	100 	\\
2005mz	&	0.017 	&	11.24 	&	0.09 	&	0.24 	&	SF	&	\nodata	&	\nodata	&	\nodata	&	\nodata	&	&	-3.24 	&	0.03 	&	0.04 	&	100 	\\
2006N	&	0.014 	&	10.82 	&	0.23 	&	0.06 	&	Pa	&	\nodata	&	\nodata	&	\nodata	&	\nodata	&	&	-3.58 	&	0.14 	&	0.12 	&	100 	\\
2006S	&	0.033 	&	10.46 	&	0.17 	&	0.11 	&	SF	&	-2.03 	&	0.33 	&	0.08 	&	6 	&	&	-1.85 	&	0.34 	&	0.21 	&	4 	\\
2006ac	&	0.024 	&	10.92 	&	0.05 	&	0.19 	&	SF	&	-1.92 	&	0.16 	&	0.05 	&	1 	&	&	-1.72 	&	0.23 	&	0.13 	&	0 	\\
2006al	&	0.069 	&	10.26 	&	0.03 	&	0.29 	&	Pa	&	-4.00 	&	99.00 	&	0.13 	&	100 	&	&	-3.74 	&	0.42 	&	0.29 	&	100 	\\
2006an	&	0.065 	&	7.54 	&	0.29 	&	0.26 	&	SF	&	-2.20 	&	99.00 	&	0.00 	&	23 	&	&	-2.80 	&	0.49 	&	0.34 	&	46 	\\
2006ar	&	0.023 	&	9.72 	&	0.05 	&	0.46 	&	SF	&	\nodata	&	\nodata	&	\nodata	&	\nodata	&	&	-1.93 	&	0.10 	&	0.07 	&	0 	\\
2006ax	&	0.018 	&	10.81 	&	0.77 	&	0.40 	&	SF	&	\nodata	&	\nodata	&	\nodata	&	\nodata	&	&	-2.76 	&	0.52 	&	0.30 	&	43 	\\
2006az	&	0.032 	&	11.28 	&	0.02 	&	0.26 	&	Pa	&	-3.43 	&	0.39 	&	0.07 	&	100 	&	&	-3.20 	&	0.21 	&	0.19 	&	96 	\\
2006bd	&	0.026 	&	11.09 	&	0.28 	&	0.09 	&	Pa	&	-4.50 	&	99.00 	&	0.12 	&	100 	&	&	-4.45 	&	0.52 	&	0.30 	&	100 	\\
2006bt	&	0.033 	&	11.09 	&	0.07 	&	0.26 	&	SF	&	-4.20 	&	99.00 	&	0.00 	&	100 	&	&	-3.50 	&	0.65 	&	0.34 	&	98 	\\
2006bw	&	0.031 	&	10.03 	&	0.08 	&	0.24 	&	Pa	&	-4.90 	&	99.00 	&	0.00 	&	100 	&	&	-3.90 	&	99.00 	&	0.00 	&	100 	\\
2006bz	&	0.028 	&	10.42 	&	0.29 	&	0.09 	&	Pa	&	-4.45 	&	0.27 	&	0.10 	&	100 	&	&	-4.41 	&	0.13 	&	0.09 	&	100 	\\
2006cf	&	0.042 	&	10.88 	&	0.10 	&	0.16 	&	SF	&	-1.77 	&	0.31 	&	0.07 	&	5 	&	&	-1.78 	&	0.36 	&	0.25 	&	3 	\\
2006cg	&	0.029 	&	9.92 	&	0.12 	&	0.27 	&	Pa	&	-4.22 	&	0.41 	&	0.07 	&	100 	&	&	-4.13 	&	0.28 	&	0.20 	&	100 	\\
2006cj	&	0.068 	&	10.42 	&	0.12 	&	0.20 	&	SF	&	-2.29 	&	0.26 	&	0.10 	&	0 	&	&	-2.26 	&	0.15 	&	0.11 	&	0 	\\
2006cp	&	0.023 	&	9.88 	&	0.06 	&	0.31 	&	SF	&	\nodata	&	\nodata	&	\nodata	&	\nodata	&	&	-3.10 	&	0.24 	&	0.16 	&	90 	\\
2006cq	&	0.049 	&	10.89 	&	0.13 	&	0.14 	&	SF	&	-2.39 	&	0.32 	&	0.08 	&	10 	&	&	-2.31 	&	0.30 	&	0.19 	&	6 	\\
2006cs	&	0.024 	&	11.09 	&	0.34 	&	0.05 	&	Pa	&	-3.79 	&	0.70 	&	0.13 	&	100 	&	&	-3.46 	&	0.42 	&	0.23 	&	99 	\\
2006ef	&	0.017 	&	10.70 	&	0.04 	&	0.08 	&	Pa	&	\nodata	&	\nodata	&	\nodata	&	\nodata	&	&	-3.84 	&	0.10 	&	0.07 	&	100 	\\
2006ej	&	0.019 	&	11.00 	&	0.24 	&	0.02 	&	Pa	&	\nodata	&	\nodata	&	\nodata	&	\nodata	&	&	-3.56 	&	0.29 	&	0.21 	&	100 	\\
2006en	&	0.031 	&	10.69 	&	0.46 	&	0.54 	&	SF	&	-1.51 	&	0.27 	&	0.07 	&	2 	&	&	-1.35 	&	0.23 	&	0.16 	&	0 	\\
2006hb	&	0.015 	&	10.95 	&	0.31 	&	0.19 	&	SF	&	\nodata	&	\nodata	&	\nodata	&	\nodata	&	&	-3.12 	&	0.51 	&	0.33 	&	80 	\\
2006kf	&	0.021 	&	10.97 	&	0.24 	&	0.09 	&	Pa	&	\nodata	&	\nodata	&	\nodata	&	\nodata	&	&	-3.12 	&	0.50 	&	0.24 	&	89 	\\
2006le	&	0.017 	&	10.19 	&	0.10 	&	0.16 	&	SF	&	\nodata	&	\nodata	&	\nodata	&	\nodata	&	&	-1.78 	&	0.52 	&	0.26 	&	9 	\\
2006mo	&	0.036 	&	10.82 	&	0.02 	&	0.17 	&	Pa	&	-3.81 	&	0.27 	&	0.04 	&	100 	&	&	-3.66 	&	0.24 	&	0.16 	&	100 	\\
2006nz	&	0.037 	&	10.62 	&	0.14 	&	0.04 	&	Pa	&	-3.77 	&	0.17 	&	0.04 	&	100 	&	&	-3.55 	&	0.11 	&	0.07 	&	100 	\\
2006oa	&	0.058 	&	8.82 	&	0.16 	&	0.13 	&	SF	&	-2.50 	&	0.20 	&	0.05 	&	7 	&	&	-2.95 	&	0.07 	&	0.08 	&	70 	\\
2006ob	&	0.058 	&	11.25 	&	0.05 	&	0.09 	&	SF	&	-2.93 	&	0.45 	&	0.09 	&	53 	&	&	-2.81 	&	0.38 	&	0.26 	&	40 	\\
2006on	&	0.068 	&	10.30 	&	0.05 	&	0.08 	&	Pa	&	-4.70 	&	99.00 	&	0.00 	&	100 	&	&	-3.27 	&	99.00 	&	0.00 	&	100 	\\
2006or	&	0.022 	&	11.07 	&	0.20 	&	0.22 	&	SF	&	\nodata	&	\nodata	&	\nodata	&	\nodata	&	&	-1.40 	&	0.22 	&	0.17 	&	0 	\\
2006os	&	0.032 	&	11.59 	&	0.93 	&	0.84 	&	SF	&	-3.47 	&	1.09 	&	0.11 	&	99 	&	&	\nodata	&	\nodata	&	\nodata	&	\nodata	\\
2006sr	&	0.023 	&	10.70 	&	0.18 	&	0.17 	&	SF	&	\nodata	&	\nodata	&	\nodata	&	\nodata	&	&	-1.86 	&	0.13 	&	0.12 	&	0 	\\
2006te	&	0.032 	&	10.31 	&	0.05 	&	0.12 	&	SF	&	-1.97 	&	0.39 	&	0.08 	&	7 	&	&	-1.86 	&	0.35 	&	0.21 	&	4 	\\
2007F	&	0.024 	&	10.06 	&	0.13 	&	0.15 	&	SF	&	-1.94 	&	0.48 	&	0.22 	&	2 	&	&	-1.89 	&	0.25 	&	0.17 	&	1 	\\
2007O	&	0.036 	&	10.70 	&	0.27 	&	0.19 	&	SF	&	-1.57 	&	0.24 	&	0.06 	&	2 	&	&	-1.52 	&	0.26 	&	0.20 	&	0 	\\
2007R	&	0.031 	&	10.98 	&	0.13 	&	0.14 	&	Pa	&	-1.66 	&	0.28 	&	0.07 	&	3 	&	&	-1.52 	&	0.27 	&	0.20 	&	2 	\\
2007S	&	0.015 	&	9.88 	&	0.36 	&	0.21 	&	SF	&	\nodata	&	\nodata	&	\nodata	&	\nodata	&	&	-1.57 	&	0.14 	&	0.11 	&	0 	\\
2007ae	&	0.064 	&	11.44 	&	0.19 	&	0.17 	&	Pa	&	-2.88 	&	0.44 	&	0.07 	&	47 	&	&	-2.32 	&	0.60 	&	0.38 	&	22 	\\
2007ar	&	0.053 	&	11.51 	&	0.04 	&	0.04 	&	Pa	&	\nodata	&	\nodata	&	\nodata	&	\nodata	&	&	-3.58 	&	0.66 	&	0.32 	&	99 	\\
2007au	&	0.020 	&	12.24 	&	0.96 	&	0.10 	&	Pa	&	\nodata	&	\nodata	&	\nodata	&	\nodata	&	&	-4.47 	&	0.53 	&	0.29 	&	100 	\\
2007ba	&	0.039 	&	11.05 	&	0.10 	&	0.18 	&	Pa	&	-3.64 	&	0.11 	&	0.03 	&	100 	&	&	-3.48 	&	0.12 	&	0.08 	&	100 	\\
2007bd	&	0.032 	&	10.76 	&	0.15 	&	0.14 	&	SF	&	-1.95 	&	0.31 	&	0.07 	&	5 	&	&	-1.81 	&	0.33 	&	0.22 	&	2 	\\
2007bz	&	0.023 	&	9.38 	&	0.11 	&	0.07 	&	SF	&	\nodata	&	\nodata	&	\nodata	&	\nodata	&	&	-1.40 	&	0.05 	&	0.04 	&	0 	\\
2007cg	&	0.034 	&	10.76 	&	0.63 	&	0.59 	&	SF	&	-2.01 	&	0.37 	&	0.09 	&	8 	&	&	-1.71 	&	0.33 	&	0.22 	&	3 	\\
2007ci	&	0.019 	&	11.13 	&	0.20 	&	0.10 	&	Pa	&	\nodata	&	\nodata	&	\nodata	&	\nodata	&	&	-3.93 	&	0.19 	&	0.14 	&	100 	\\
2008L	&	0.019 	&	10.13 	&	0.09 	&	0.55 	&	Pa	&	\nodata	&	\nodata	&	\nodata	&	\nodata	&	&	-4.26 	&	0.48 	&	0.19 	&	100 	\\
2008af	&	0.034 	&	11.48 	&	0.09 	&	0.17 	&	Pa	&	-4.37 	&	99.00 	&	0.57 	&	100 	&	&	-3.27 	&	0.62 	&	0.30 	&	92 	\\
2008bf	&	0.026 	&	11.39 	&	0.28 	&	0.03 	&	Pa	&	-4.30 	&	99.00 	&	0.00 	&	100 	&	&	-3.43 	&	99.00 	&	0.00 	&	100 	\\
\enddata
\end{deluxetable}

\clearpage

\begin{deluxetable}{lcccccc}
\tabletypesize{\scriptsize}
\tablewidth{0pt}
\tablecaption{\label{tab:hostresults} The weighted-mean and r.m.s. scatter of SN Ia Hubble residuals in different host environments}
\tablehead{
\colhead{Group} & \colhead{} & \colhead{$N$} & \colhead{Mean residual} & \colhead{Error} & \colhead{r.m.s.} & \colhead{Error} \\
\cline{3-7}
\colhead{} & \colhead{} & \colhead{} & \colhead{(mag)} & \colhead{(mag)} & \colhead{(mag)} & \colhead{(mag)}
}
\startdata
Locally Passive      & & 194 & -0.043 & 0.013 & 0.180 & 0.009  \\
Locally Star-Forming & & 174 &  0.038 & 0.013 & 0.172 & 0.009  \\ \hline
Diff.                & &     & \textbf {0.081} & \textbf {0.018} & &  \\ \hline
\colhead{} & \colhead{} & \colhead{} & \colhead{} & \colhead{} & \colhead{} & \colhead{} \\ \hline
Globally Passive      & & 194 & -0.043 & 0.013 & 0.180 & 0.009  \\
Globally Star-Forming & & 455 &  0.006 & 0.008 & 0.167 & 0.006 \\ \hline
Diff.                 & & & \textbf {0.049} & \textbf {0.015} & &  \\ \hline
\colhead{} & \colhead{} & \colhead{} & \colhead{} & \colhead{} & \colhead{} & \colhead{} \\ \hline
High-Mass ($\log \mstellar>10$) \ & & 464 & -0.022 & 0.008 & 0.172 & 0.006  \\
Low-Mass ($\log \mstellar\leq10$) & & 184 &  0.035 & 0.012 & 0.164 & 0.009  \\ \hline
Diff.                             & &     & \textbf {0.057} & \textbf {0.014} & &  \\ \hline
\colhead{} & \colhead{} & \colhead{} & \colhead{} & \colhead{} & \colhead{} & \colhead{} \\ \hline
For comparison          & &      &       &       &       &       \\ \hline
YONSEI Cosmology        & & 941  &  0.002 & 0.006 & 0.196 & 0.005 \\
Full Local sample       & & 368  & -0.003 & 0.009 & 0.175 & 0.006 \\
\enddata
\end{deluxetable}

\clearpage

\begin{deluxetable}{lcccccccc}
\tabletypesize{\scriptsize}
\tablewidth{0pt}
\tablecaption{\label{tab:cosmology} The best-fit flat $\Lambda$CDM parameters}
\tablehead{
\colhead{Group} & \colhead{} & \colhead{SNe} & \colhead{$\Omega_{M}$} & \colhead{$\alpha$} & \colhead{$\beta$} & \colhead{$M_{B}$} & \colhead{$\sigma_{int}$}  & \colhead{$\chi^{2}$/D.O.F.}
}
\startdata
Locally Passive & & 194 & $0.33^{+0.09}_{-0.10}$ & $0.18^{+0.03}_{-0.02}$ & $2.96^{+0.31}_{-0.29}$ & $-19.12^{+0.04}_{-0.05}$ & 0.104 & $190.62/190$ \\
Locally Star-Forming & & 174 & $0.31\pm0.09$ & $0.14^{+0.04}_{-0.03}$ & $3.40^{+0.40}_{-0.37}$ & $-19.02\pm0.05$ & 0.111 & $170.58/170$ \\ \hline
Full Local sample       & & 368  & $0.29^{+0.07}_{-0.05}$ & $0.14^{+0.02}_{-0.01}$ & $3.16^{+0.25}_{-0.24}$ & $-19.07\pm0.03$ & $0.117$ & $363.60/364$   \\
YONSEI Cosmology        & & 941 & $0.31^{+0.04}_{-0.03}$ & $0.15^{+0.00}_{-0.01}$ & $3.08^{+0.16}_{-0.15}$ & $-19.06^{+0.02}_{-0.01}$ & 0.135 & $ 935.36/937$ \\
\enddata
\end{deluxetable}

\clearpage

\end{document}